# Quadrimechanica
*Motion: observation and detection*

*Roberto Assumpção*
*PUC-Minas, Av. Pe. Francis C. Cox, Poços de Caldas- MG  37701-355, Brasil*
assumpcao@pucpcaldas.br

# # - Abstract

Classical, Quantum and Relativistic mechanics elect time and space as fundamentals, extracting the measure of motion –velocity– from this static space-time platform. Conversely, the timelessness of Statistical mechanics computes the physical flow as an evolution of (dynamical) states which, at the end, comes to be what would be denoted as time. Moreover, the general structure of mechanics is such that the notion of observability is mixed up with that of measurability. In such a context, ignoring the distinction between "observable entities" and "measurable quantities", and projecting dynamical objects on a static background, the measured value of physical entities is taken as the true or real value of the entity. This contribution emphasises this distinction, correlating the concepts of  "real values" and "experimental values" and indicating that motion is always underestimated in the spatiotemporal platform; following a gauge transformation, an argument centred on *measurability* favours a relational "non-relativistic" formulation, with potentially analogous predictions.

# # 1 – Introduction

At the present time it is probably much more simple to develop a measurement process to any physical entity than to specify the physical nature of the obtained quantity; the proliferation of processes, methods and compatible instrumentation was not accompanied by a unifying theoretical perspective. In fact, a well-recognised situation [1] is that Nature appears divided in to " theories of the Macro " and " theories of the Micro ", not to say "physics of the slow" and "physics of the swift".

Macroscopic phenomena, large–scale objects, are governed by relativistic mechanics that, in some sense, incorporates classical mechanics. Microscopic objects follow the laws of quantum mechanics. Multiple-small-scale-objects are assumed to be governed by statistical mechanics, in collective (complex) phenomena. A "swift–slow" scheme of classification could tell the same story, with a more sounded physical basis once the pragmatism of motion appears more satisfactory than that of size, except for quantum mechanics.

The concepts of big and small are consistently treated by quantum physics who gave an absolute meaning to the size as well as relativity theory gave an absolute meaning to the motion. Throughout the proposal of the postulate of relativity, in the context of the Michelson-Morley experiment, from which the Lorentz-Fitzgerald contraction can be deduced, Einstein [2] pointed out that if nature conspires in such a way to prevent us from determining experimentally the absolute motion of our instruments, then the notion of an absolute velocity has no meaning.

Analogously, revising the classical ideas of Causality, Dirac [3] pointed out that in order to give an absolute meaning to size, such as is required for any theory of the ultimate structure of matter, it is necessary to assume that there is a limit of the power of observation: if the object under observation is such that the unavoidable limiting disturbance is negligible, then the object is big in the absolute sense; if, on the other hand, the limiting disturbance is not negligible, then the object is small in the absolute sense.

Remarkably, these diverse approaches to the absolute coincide on accepting the fact that motion detection, a difficult experimental measurement even in Classical terms, can give rise to an absolute value, precisely in the case of a small and fast object. Velocity measurements imply detection of two events evolving on distinct space locations; conversely, single point detection associated to motion transfer (energy / momentum) has the advantage of a direct measurement but is experimentally limited by the uncertainty principle at the quantum level and by the 'motion transfer model' in any situation. The figure below illustrates a conventional indirect velocity measurement in the space – time platform.

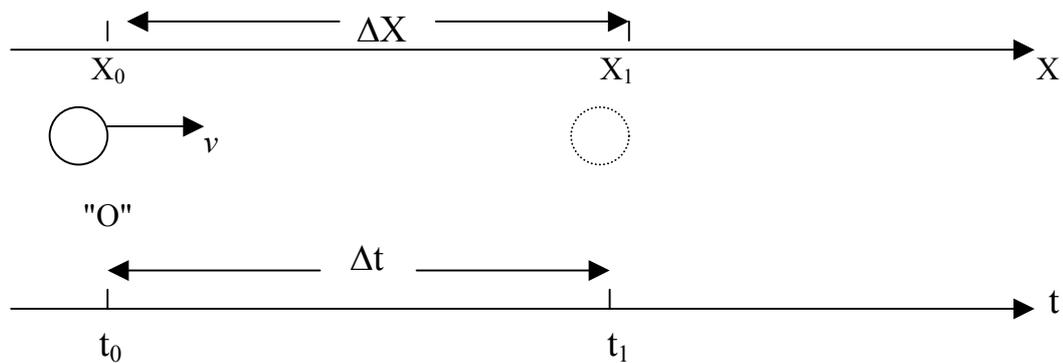

**Figure 1** - Velocity measurement in space-time: an observer + instrumentation "O" at $X_0$ observes the passage of an object (velocity $v$) and realises a measurement of the time lasted for the dislocation between the positions $X_0$ and $X_1$.

Now, even adopting the space-time platform, motion detection should be invariant in a sense that the physical observable entity (velocity) corresponds to the concrete measured quantity (the particular velocity), independent of the coordinates of the platform. This implies independent measurements on the platform and a way of vanishing the background coordinates (time <u>and</u> space); another possibility is to balance with an off–platform gauge. The analysis here follow this "off background" method.

## # 2 – *Time Measurements*

The time $\Delta t_m$ measured by the observer "O" for the dislocation of the object between the positions $X_0$ and $X_1$ is given by the relation:

$$\Delta t_m = \Delta t + \Delta t_i \quad (1)$$

Where $\Delta t$ is the effective time that the particle lasts between $X_0$ and $X_1$, $\Delta t_m$ the measured time of the motion between $X_0$ and $X_1$ and $\Delta t_i$ the time required for the signal (object's image) to come back to the observer "O" at $X_0$.

Equation (1) is an uncertainty relation, that is:

$$\Delta t \leq \Delta t_m$$

This means that the real time $\Delta t$ is not exactly equal to the time $\Delta t_m$ detected by instrumentation. Thus, relative to the phenomena, Observer is always in the past, once the detected value of the time is limited by the time required for the signal to arrive at measurement location. The real time of motion can be written as:

$$\Delta t = \frac{\Delta x}{v} \qquad (2)$$

where $\Delta X$ is the $X_0 X_1$ distance and $v$ the (real) velocity of the object.

The signal (transportation) time $\Delta t_i$ can be written as,

$$\Delta t_i = \frac{\Delta x}{v_s} \qquad (3)$$

where $v_s$ is the effective velocity of the signal carrier.

From equations (2) and (3) the time $\Delta t_i$ can be written as a function of $\Delta t$, that is:

$$\Delta t_i = \frac{v}{v_s} \Delta t \qquad (4)$$

By means of this relation, equation (1) becomes:

$$\Delta t_m = \Delta t \left(1 + \frac{v}{v_s}\right) \qquad (5)$$

This equation relates the real time $\Delta t$ of a fact (motion of an object) with the measured $\Delta t_m$ time of the observed effect (passage of the object throughout space from $X_0$ to $X_1$). This means that the detectable effect of a real fact (evolving at velocity $v$ and lasting $\Delta t$) is delayed by a factor $\{1 + v / v_s\}$. Thus, for any finite signal velocity, a time measurement overestimates the real value. Moreover, since all measured times can be computed by this same "out of platform" factor, an independent determination of the quotient $v / v_s$ can settle relation (5), that is, time, into concrete physical grounds; this demands for an analysis of velocity measurements.

## # 3 – *Velocity Measurements*

The velocity $v_m$ measured by the observer "O" at $X_0$ is:

$$v_m = \frac{\Delta x}{\Delta t_m} \qquad (6)$$

Employing relations (2) and (5) this equation becomes:

$$v_m = \frac{v}{1 + \dfrac{v}{v_s}} \quad (7)$$

This equation relates the real velocity $v$ of a fact with the measured velocity $v_m$ of the effect. Note that the real velocity value is different from the measured one by a factor $\{1 + v / v_s\}$. Thus, for any finite $v_s$, $v_m < v$, that is, a velocity measurement underestimates the true value.

These equations show that time and velocity measurements are governed by the same $\{1 + v / v_s\}$ factor, where the quotient $v / v_s$ determines the distance between the experimental and the true values. In order to increase precision on both measurements, it is convenient to reduce the time $\Delta t_i$ required for signal transportation, that is, increase $v_s$, employing a fast signal carrier, such as a light pulse. This demands for knowledge of the effective value of the velocity of light.

The measured value of the propagation of light in vacuum is '$c$'. Employing relation (7) with $v = v_s$ and taking the experimental value $v_m = c$, the value of the light pulse is:

$$v_s = 2c \quad (8)$$

Substituting this value in relations (5) and (7):

$$\Delta t = \left(1 - \frac{v_m}{2c}\right)\Delta t_m \quad (9)$$

$$v = \frac{v_m}{\left(1 - \dfrac{v_m}{2c}\right)} \quad (10)$$

Equations (9) and (10) show explicitly the distinction between the observable value (the intangible real value of times and velocities) and the measured values of these entities; both equations show that the off platform factor $\{1 - v_m / 2c\}$ governs the measurement. However, the action of this factor is distinct: in (9) the factor simply corrects the experimental value, reflecting the well known relativistic dilation effect; conversely, equation (10) can be worked out to show that:

$$\frac{1}{v_m} = \frac{1}{v} + \frac{1}{2c}$$

This last form shows that the measured velocity $v_m$ is harmonically related to the real value $v$ by means of the signal velocity ($2c$ or $v_s$); that is, the measurement process collapses the true (observable) value to the experimental (measurable) data.

# 4 – Comments

The distinction between Observability and Measurability is anccient; in the words of Plato [4], " *the starry heaven which we behold is wrought upon a visible ground, and therefore, although the fairest and most perfect of visible things, must necessarily be deemed inferior far too the true motions of absolute swiftness and absolute slowness, which are relative to each other, and carry with them that which is contained in them, in the true number and in every true figure. Now, these are to be apprehended by reason, but not by sight* ". This solves the enigma of intangibility.

There is an unavoidable intangibility in the calculation of measurable data, the theoretical method enabling only the possibility (not the probability) of obtaining an <u>exact</u> (quantum mechanical sense) result. Therefore, the presence of the uncertainty principle in equation (1) is quite reasonable; moreover, (1) and (10) can also be considered as gauge transformations:

$$\Delta t_m \to \Delta t + \Delta t_i \quad (1.A)$$

$$\frac{1}{v_m} \to \frac{1}{v} + \frac{1}{v_s} \quad (10.A)$$

In technical terms, the gauge transformation (1.A) leads to the *velocity gauge* (10.A). A priori, any choice of the gauge is possible, once both satisfy the condition of measurability and will 'act' as the same factor in equations (9) and (10). On the other hand, absence of this gauge transformation means no gauge at all.

The advantage of the $v_s$ gauge is the substitution of the static space–time platform by a velocity–space dynamical background, where the physical flow do happens, in which relative localisation of physical entities is the rule and from which a spatiotemporal stage can be extracted.

Finally, there seems to be a fundamental relationship between velocity and space but not among time and those; equation (9) sets time in to the concrete basis of a balance of velocities but it also tells that time depends on the existence of the physical flow. This is consistent with the statistical mechanics (thermodynamically sense) notion that considers this entity as state dependent, appearing only connected to the evolution of states (motion).

## *# 5 – Conclusions*

Observability plays a central hole in physics, leaving the choice of the background to describe the physical flow in a second plane; however; the velocity–space dynamical platform appears superior to the static space-time stage once localisation is intrinsically relative (relational).

A partial equivalence between the spatiotemporal and the velocity–space platform can be achieved by a gauge transformation, resulting on the indication that motion is always underestimated in space–time.

Development of a formalism based on the velocity gauge may give rise to a synthetic treatment of the physical flow in terms of suitable coordinates; this contribution also emphasises that the *velocity gauge* approach may retain the basic ingredients of classical, quantum and statistical mechanics, without the limitations imposed by relativity.

# # 6 – *References*

# # 7 – *Acknowledgements*

The original '*gauge transformation*' is due to A Teixeira Neto; discussions with A C França, A J Roberto Jr., L F Delboni, M M H Barreira and W Amstalden are gratefully acknowledged. Comments/suggestions are welcomed and can also be sent to assump@fem.unicamp.br.